\documentclass{elsart}
\usepackage{natbib}
\usepackage{graphicx}
\usepackage{epsfig}
\usepackage{amssymb}
\begin{document}

\begin{frontmatter}

\title{Cluster scaling and its redshift evolution from XMM-Newton.
\thanksref{titlefn}}
\thanks[titlefn]{Based on observations obtained with {\em XMM-Newton}, an ESA science mission with instruments and contributions directly funded by ESA
Member States and the USA (NASA).}
\author[MP]{A. Finoguenov},
\author[MP]{H. B\"ohringer},
\author[Bh]{J. P. F. Osmond},
\author[Bh]{T. J. Ponman},
\author[UIUC]{A.J.R. Sanderson},
\author[MP]{Y.-Y. Zhang},
\author[MP]{M.Zimer}
\address[MP]{ Max-Planck-Institut f\"ur extraterrestrische Physik,
  Giessenbachstra\ss e,  PF1312, 85748 Garching, Germany}
%\address[UM]{Joint Center for Astrophysics, Physics Department, University of
%  Maryland, Baltimore County, Baltimore, MD 21250, USA}
%\address[CI]{Observatories of the Carnegie Institution of Washington, 813
%Santa Barbara Street, Pasadena, CA 91101, USA}
\address[Bh]{School of Physics \& Astronomy, University of Birmingham,
  Edgbaston, Birmingham B15 2TT, UK}
\address[UIUC]{ Department of Astronomy, University of Illinois,
        1002 West Green Street, Urbana, IL 61801, USA}
\begin{abstract}
We put together the results of XMM-Newton observations of a number of
representative group and cluster samples at low and high redshifts. These
results confirm the entropy ramp as an explanation of the observed scaling
relations. We observe a mild evolution in the entropy of clusters. The
observed degree of evolution is consistent with expectations of the shock
heating at a fixed overdensity (500) with respect to the critical density in
$\Lambda CDM$. The study of the evolution in the pressure scaling imposes
strong requirements in the definition of the average temperature of the
cluster. The scaling temperature should be consistent to better than the
10\% level. Once such a consistency is achieved, no additional evolution in
the pressure has been detected in addition to the prediction of the shock
heating in the $\Lambda CDM$ Universe.
\end{abstract}


\end{frontmatter}

\section{Introduction}

Comparative studies of the scaling relations in clusters of galaxies reveal
strong deviations of the observed relations from predictions based on
self-similar collapse, e.g. the observations show a steeper $L_x-T$
relation than predicted by the self-similar laws (Kaiser 1986).
These deviations are thought to be best characterized by the
injection of energy (preheating) into the gas before clusters collapse
(Kaiser 1991; Evrard \& Henry 1991).  
Recently, an analysis of a large compilation of entropy profiles
on groups and clusters of galaxies also
required at $r_{500}$ much larger entropy levels than was thought before
(Finoguenov et al. 2002) and modifying the concept of the entropy floor to
the entropy ramp at $0.1r_{200}$ (Ponman et al. 2003). Reproduction of these
results both analytically and numerically, strongly supports the scenario of
Dos Santos \& Dore (2002), where an initial adiabatic state of the infalling
gas is further modified by the accretion shock (Voit \& Ponman 2003).  As a
supporting evidence to the latter, Ponman et al. (2003) noticed a
self-similarity in the entropy profiles, once scaled to $T^{0.65}$. Some
XMM-Newton observations are consistent with this result (Pratt \& Arnaud
2003). A major change introduced by these studies is that groups of galaxies
can again be viewed as scaled-down versions of clusters, yet the scaling
itself is modified. Other evidence for the departure of groups from the
trends seen in clusters, such as the slope of the $L-T$ relation, has been
recently refuted by Osmond \& Ponman (2004).

The idea of this {\it contribution} is to check the consistency between the
data and both the concept and the level of the modified entropy scaling.
While we give an overview of the results here, the details of the data
analysis could be found in Zhang et al. (2004); Finoguenov et al. (2004b,
and in prep.).

For our study we have selected 14 groups in Mulchaey et al. (2003) in the
redshift range $0.012<z<0.024$ with publicly available XMM-Newton (Jansen et
al. 2001) observations.  Most of the groups in the Mulchaey et al. (2003)
sample were found by cross-correlating the ROSAT observation log with the
positions of optically-selected groups. Their final group list contains 109
systems.

The REFLEX-DXL galaxy cluster sample, comprising {\bf d}istant {\bf X}-ray
{\bf l}uminous objects within REFLEX, was constructed from the REFLEX galaxy
cluster survey covering the ROSAT detected galaxy clusters above a flux
limit of $3 \cdot 10^{-12}$ erg s$^{-1}$ cm$^{-2}$ in the 0.1 to 2.4 keV
band in an 4.24 ster region of the southern sky (see B\"ohringer et al. 2004
for details). The REFLEX-DXL clusters form a volume limited subset of REFLEX
in the redshift range 0.27 to 0.31 including 14 members. The properties of
the REFLEX-DXL clusters are described in Zhang et al. (2004).

In addition to these two samples, we have added some of the published work
on the nearby clusters, A754 (Henry et al. 2004), A3667 (Briel et al. 2004),
A3562 (Finoguenov et al. 2004a); A478 (Sanderson et al. 2004); A3558
(Rossetti et al. in prep.); A3266 (Finoguenov et al. 2005). Addition of
these data allows us to demonstrate to which extent the scaling works at low
redshift.

At the moment a number of prescriptions exists on how to scale the cluster
properties and look for their evolution (e.g. Voit 2004). We choose the
following approach to scale the observables. The calculation of the
$r_{500}$ is the following: $r_{500}=0.45 {\rm Mpc} \times
\sqrt{kT/{\rm keV}} h_{70}^{-1} h(z)^{-1}$, where the scaling in
Finoguenov et al. (2001) for $h_{50}$ ($r_{500}\approx 0.63 {\rm
Mpc}\sqrt{kT/{\rm keV}}$) is translated into our assumption for
$h_{70}=1$. We use \mbox{$h(z)=(\Omega_M (1+z)^3 + \Omega_\Lambda)^{1/2}$},
suitable for our choice of cosmological model. In Finoguenov et al. (2001)
it has been demonstrated that the cosmological corrections are negligible in
deriving the scaling for $r_{500}$ in their sample of local clusters. These
corrections are, however important for REFLEX-DXL.

% revise this paragraph
The normalization of the empirical entropy scaling is taken from Ponman et
al. (2003; hereafter PSF) and rescaled for the difference in the assumption
for the Hubble constant.

\begin{figure}
\includegraphics*[width=10.cm]{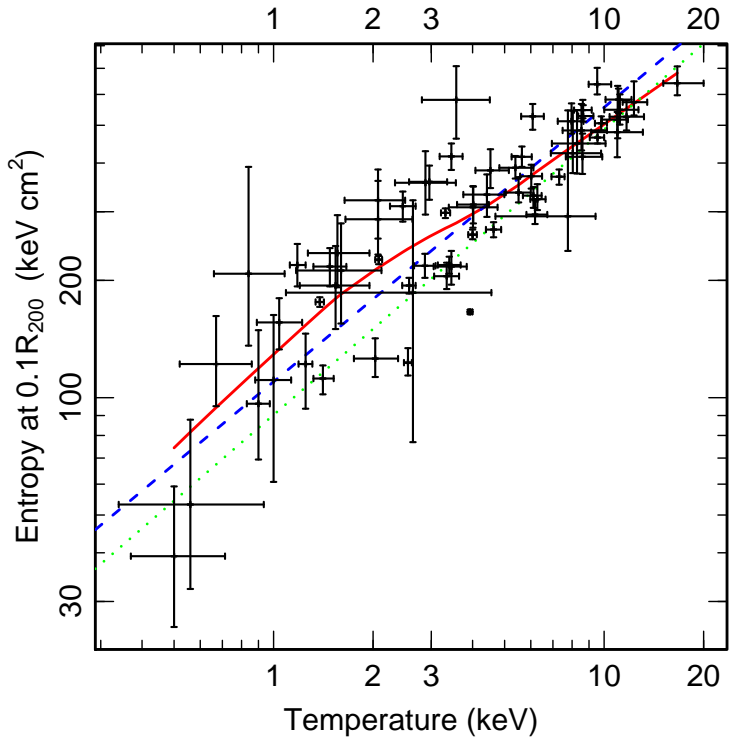}
\caption{Reanalysis of the entropy ramp data of Ponman et al. (2003). Blue
  dashed line represents the best-fit unweighted power law (slope is equal
  to $0.70\pm0.05$). The green dotted line stands for the weighted
  orthogonal regression (slope equals to $0.73\pm0.07$). The red solid line,
  shows the results of the locally-weighted pseudo-non-parametric
  analysis. This analysis reveals a glitch in the $S-T$ behavior at 4 keV
  temperature.
\label{fin0}}
\end{figure}

A suggested modified entropy scaling of PSF reads $S \sim T^{0.65}
h(z)^{-4/3}$. We also take into account that to fit the clusters hotter than
5 keV, the normalization of this relation should be 20\% lower.  In
Fig.\ref{fin0} we illustrate this issue by performing the locally-weighted
pseudo-non-parametric analysis (for further insights and references, see
Sanderson et al. 2004). While this is done here to properly reproduce the
mean entropy of the clusters, observed previously, a glitch in the $S-T$
(and $P-T$) relation implied by such an approach, will be studied
elsewhere. In the following we will use the entropies measured at $r_{500}$
for a 10 keV systems and apply scaling with $({kT \over 10 {\rm
keV}})^{0.65}$. In addition to the scaling of the entropy and a fixed
fraction of virial radius, we adopt a radial behavior of the entropy as
$r^{1.1}$, which for the local sample has been shown to work outside the
$0.1r_{500}$ (Pratt et al. 2003).

In the analysis of clusters, we will also present the scaled pressure plots.
As entropy, $T n^{-2/3}$, scales as $T_w^{2/3}$ (where $T_w$ is the weighted
temperature), the density scales as $T_w^{1/2}$ and the pressure $T n \sim
T_w^{3/2}$. For high-redshift clusters, we introduce a correction for the
evolution of the critical density, which is proportional to
$h(z)^2$. Introduction of the correction for the evolution of the critical
density to either pressure or entropy is appropriate only if the shock
heating of the accreted gas is the dominant mechanism defining the
thermodynamics of the ICM and the resulting agreement of the data with the
local scaling confirms the major role of the shock heating in establishing
the entropy and pressure profiles for clusters. A sample of high-redshift
groups would be needed to extend this conclusion to groups.

As we are not aware of any prescription for the pressure, in this {\it
contribution} we use the fit to A478 data and discuss whether it is
representative. Including the prescription for the scaling of the pressure
with the average cluster temperature, the pressure model, implied by the
A478 data reads as $P=10^{-9} \cdot (1+({r \over 0.05 r_{500}})^2)^{-1} ({kT
\over 10 {\rm keV}})^{1.5}$ ergs cm$^{-3}$. In addition, we adopt a factor
of 1.5 lower pressure normalization for systems with temperatures less than
5 keV in concordance with the entropy scaling.

The pressure within the central $0.2 r_{500}$ in clusters with on-going
mergers (most remarkably A3558 and A3266) lies below our pressure model.
The same clusters also reveal an entropy which is much higher in the center,
which may at least partly explain the effect. At radii beyond $0.2r_{500}$,
there is less scatter among the clusters and the adopted pressure model
performs well.

\begin{figure}
\includegraphics*[width=7.cm]{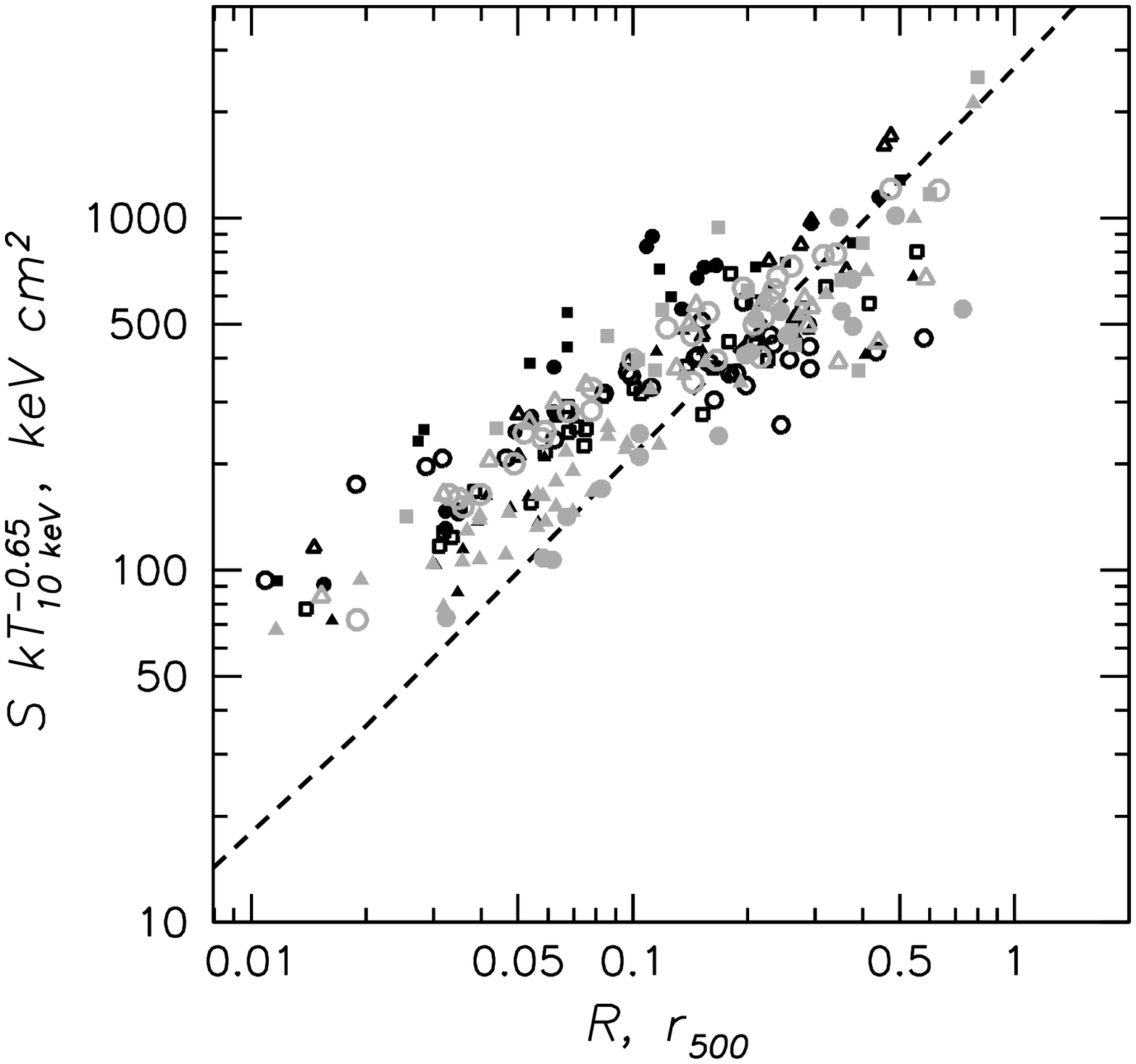}
\includegraphics*[width=7.cm]{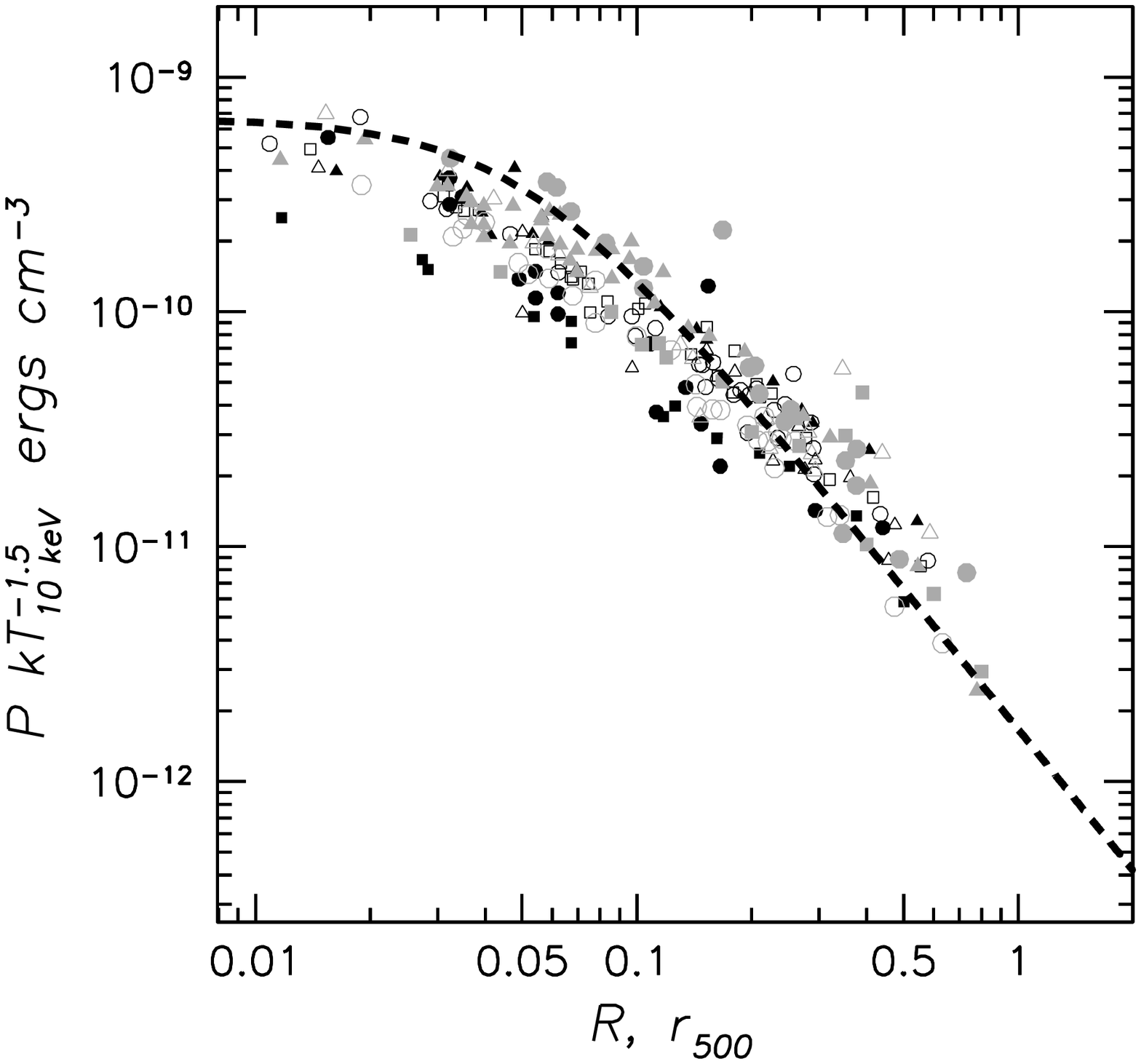}\\
\includegraphics*[width=7.cm]{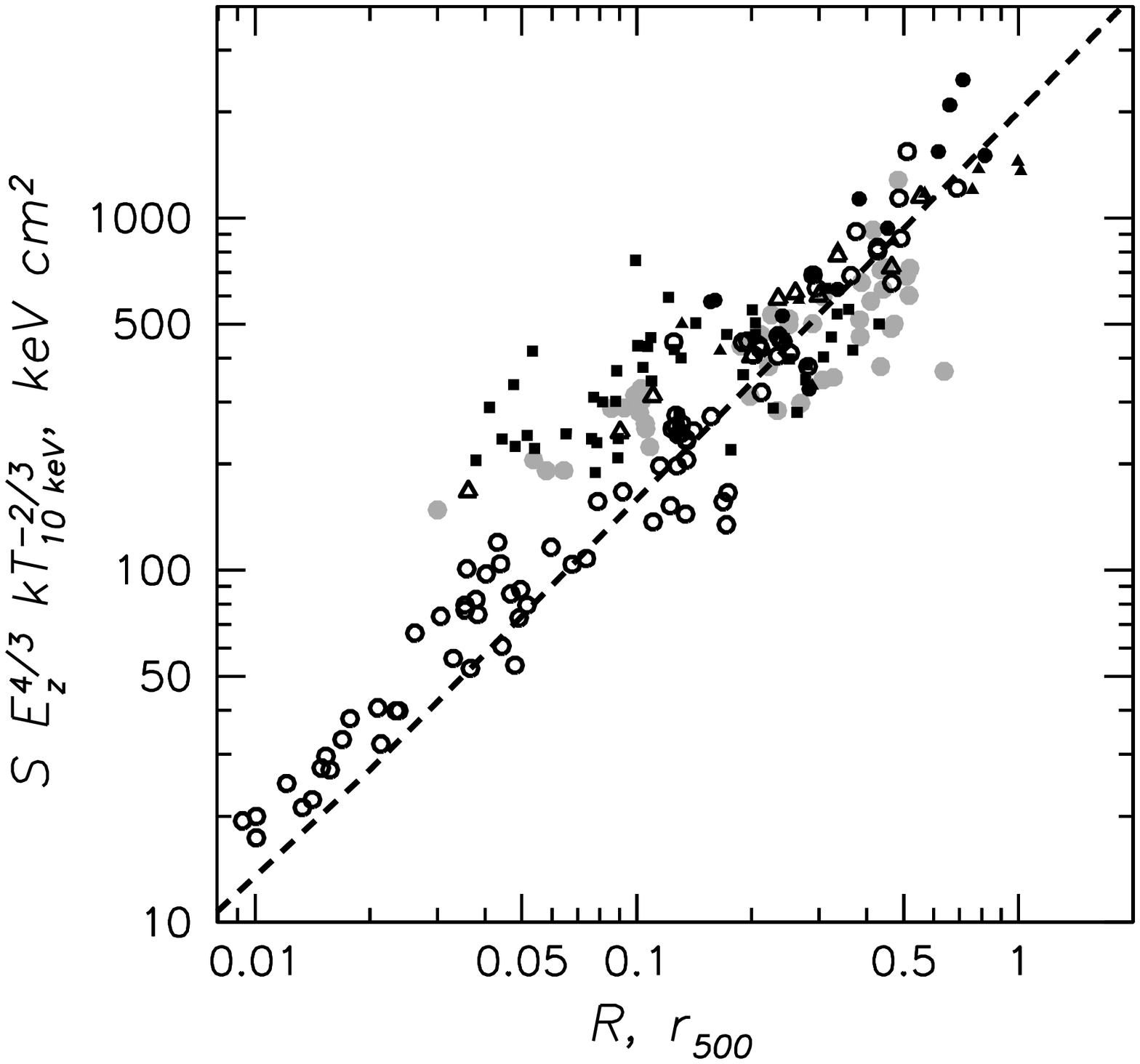}
\includegraphics*[width=7.cm]{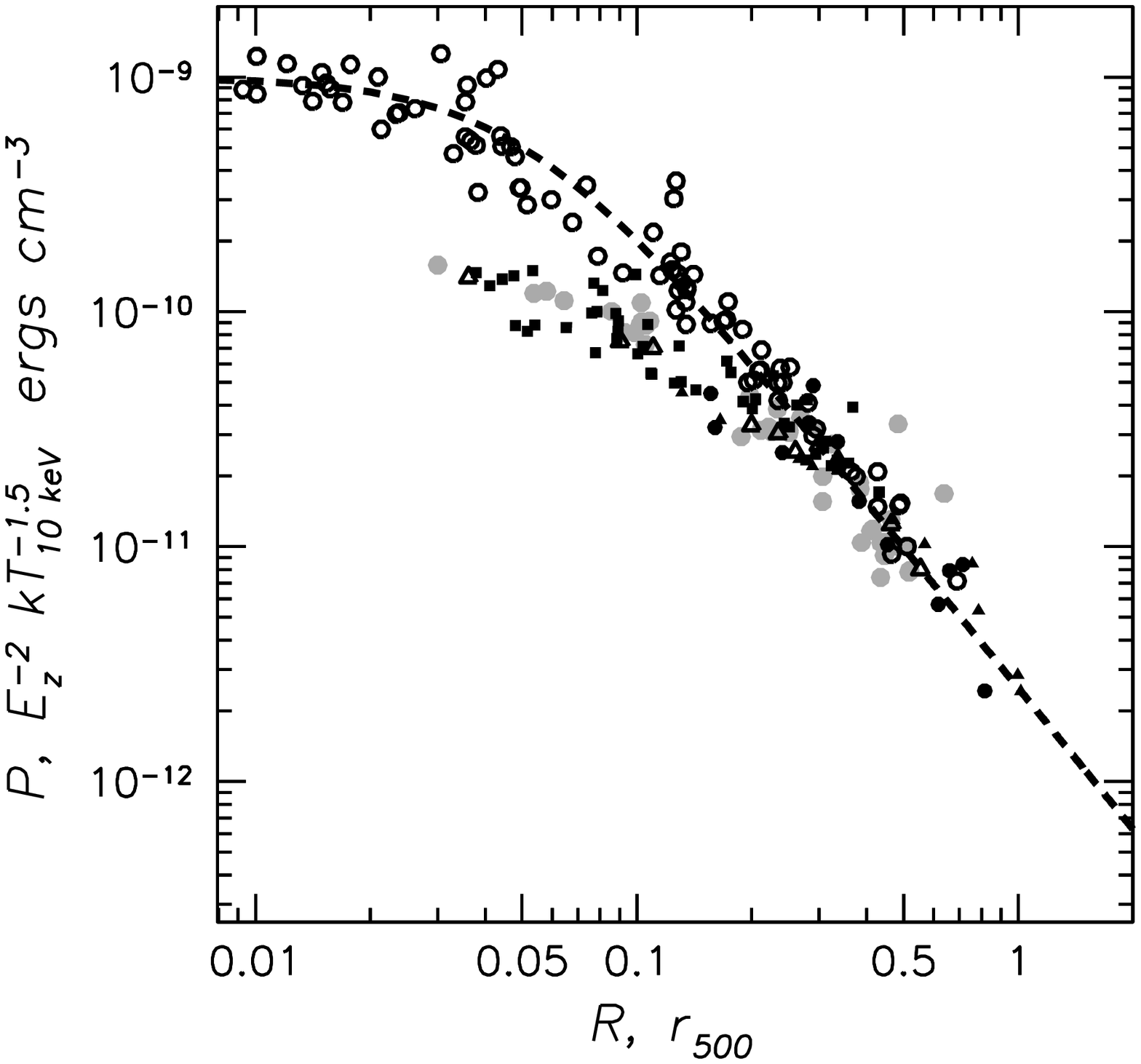}\\
\includegraphics*[width=7.cm]{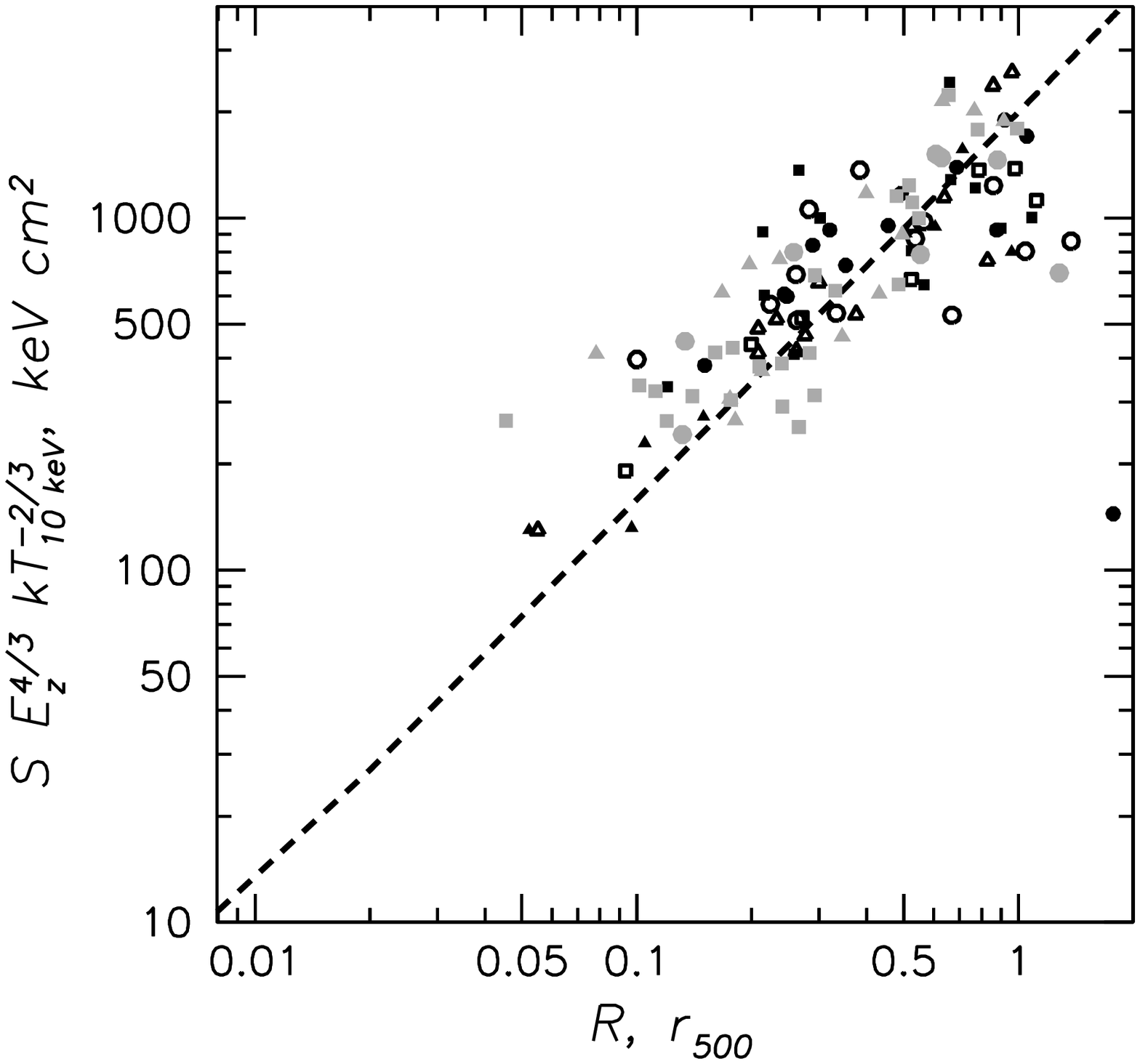}
\includegraphics*[width=7.cm]{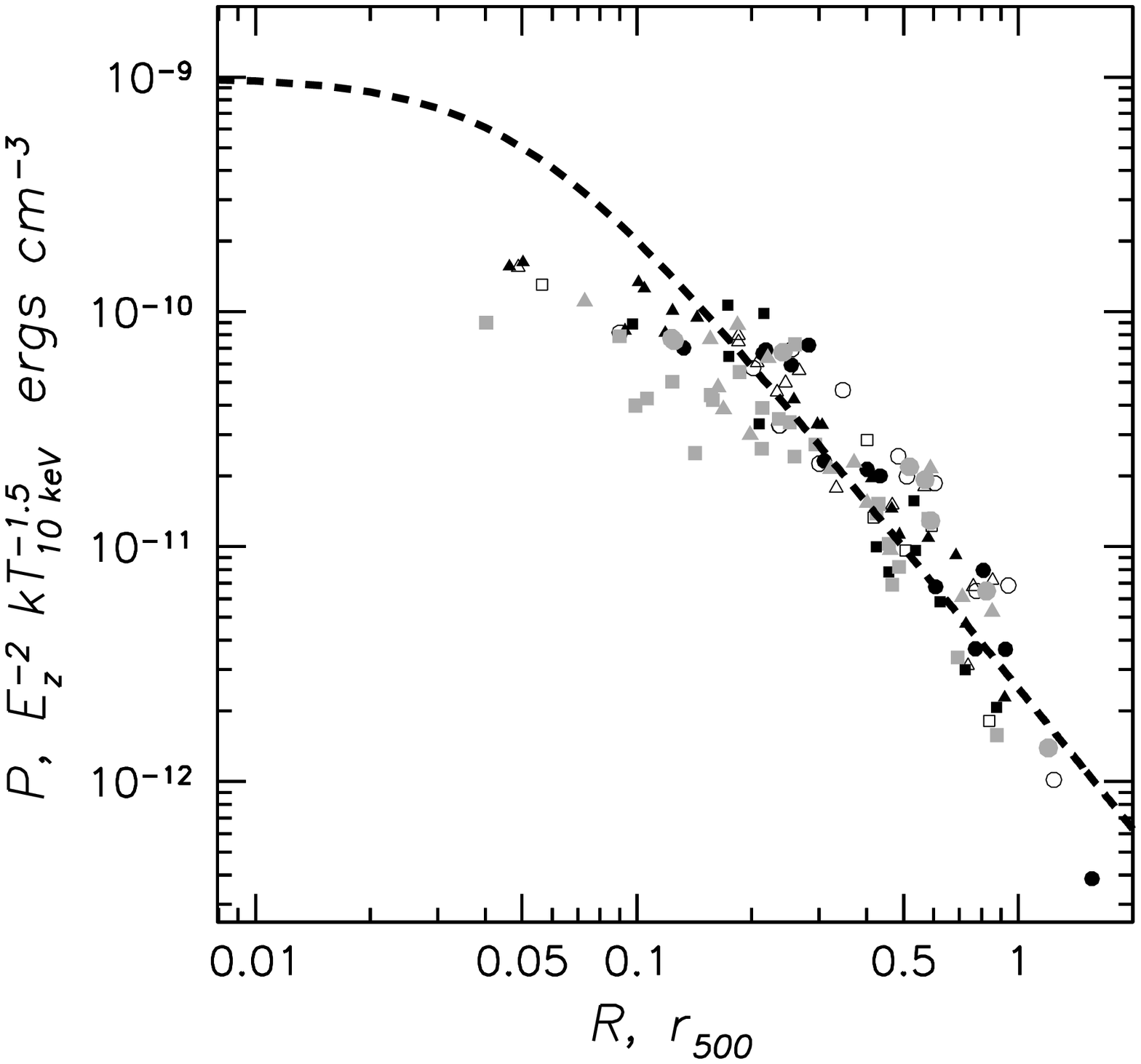}

\caption{Scaled entropy (left) and pressure (right) profiles. Upper row
  displays the results for the nearby groups ($0.012<z<0.024$). Middle row
  shows nearby hot ($kT>5$ keV) clusters. Lower row presents distant X-ray
  luminous clusters ($0.27<z<0.31$). The systems are separated by using
  different symbols. The dashed lines in the left column of panels show the
  entropy profile of $r^{1.1}$, normalized to the results of Ponman et
  al. (2003) at $0.1 r_{200}$.  The dashed lines in the right column of
  panels show our model for the pressure profile, as described in the text.
\label{fin1}}
\end{figure}

Our main results are presented in Fig.\ref{fin1} and imply:

(1) Strong deviations from the $r^{1.1}$ law are occurring inside
$0.2r_{500}$ for groups, while in regular clusters such deviations are
generally confined to $0.1 r_{500}$ (Pratt \& Arnaud 2003). It may indicate
that the effect of non-gravitational processes is important to a larger
extent in the groups. However, similar level of deviations is also present
in the merger clusters.

(2) At radii $0.2<r/r_{500}<0.6$, the entropy profiles in the groups are on
average flatter than the $r^{1.1}$ law. This result raises an interest in a
systematic study of entropy profiles at $r/r_{500}>0.6$. The level of the
observed flattening is around 100 keV cm$^2$ and could be related to the
cooling threshold, discussed in Voit et al. (2003). At the same radii, the
pressure profile in groups is on average flatter than in clusters, which
is seen as on average higher pressure at outskirts of the groups compared to
the model.

(3) A comparison between the prediction of the entropy according to the
evolution of the shock heating in the $\Lambda CDM$ Universe and the data
can explain the entropy of the gas. The radial behavior of the entropy is
flatter, compared to the index of $1.1-1.2$ predicted in a hierarchical
cluster growth with no feedback effects (Tozzi \& Norman 2001; Voit
2004). This result implies that either the growth of clusters has been
slower or feedback effects are significant. Slower accretion rates support a
suggestion of a dark energy dominated Universe (Schuecker et al. 2003).

(4) At high redshift, the typical pressure of the gas is found to be
in agreement with prediction of the evolution, once a consistent
definition of the mean temperature has been assumed for scaling. 
The effect of energy band in determining the mean temperatures has
been discussed in Zhang et al. (2004) in application to the REFLEX-DXL
sample and by Mazzotta et al (2004) in application to clusters in general.
% 
% Merger activity could increase the pressure at
% the initial phases. Or is it our estimate of the temperature for the scaling
% radius that does not work? We could use mass for scaling...

\end{document}